\documentclass{ws-procs9x6}

\begin{document}

\title{The Strangeness Physics Program at CLAS}

\author{D. ~S. CARMAN}

\address{Department of Physics, Ohio University \\
Athens, OH 45701, USA\\ 
E-mail: carman@ohio.edu\\
(for the CLAS Collaboration)}

\maketitle

\abstracts{An extensive program of strange particle production off the 
proton is currently underway with the CEBAF Large Acceptance Spectrometer 
(CLAS) in Hall B at Jefferson Laboratory. This talk will emphasize strangeness 
photo- and electroproduction in the baryon resonance region between 
$W$ =1.6 and 2.5~GeV, where indications of $s$-channel structure are 
suggestive of high-mass baryon resonances coupling to kaons and hyperons 
in the final state.  Precision measurements of cross sections and
polarization observables are being carried out with both electron 
and real photon beams, both of which are available with high polarization
at energies up to 6~GeV.} 

\section{Introduction}

A key to understanding the structure of the nucleon is to understand
its spectrum of excited states.  However understanding nucleon resonance
excitation provides a serious challenge to hadronic physics due to the
non-perturbative nature of QCD at these energies.  Recent symmetric quark 
model calculations predict more states than have been seen 
experimentally\cite{capstick}.  Mapping out the spectrum of these excited 
states will provide for insight into the underlying degrees of freedom of 
the nucleon.

Most of our present knowledge of baryon resonances comes from reactions
involving pions in the initial and/or final states.  A possible explanation
for the so-called missing resonance problem could be that pionic coupling
to the intermediate $N^*$ or $\Delta^*$ states might be weak.  This
suggests a search for these hadronic states in strangeness production
reactions.  Beyond different coupling constants (e.g. $g_{KNY}$ vs. 
$g_{\pi NN}$), the study of the exclusive production of $K^+\Lambda$ and 
$K^+\Sigma^0$ final states has other advantages in the search for missing 
resonances.  The higher masses of the kaon and hyperons, compared to their 
non-strange counterparts, kinematically favor a two-body decay mode for 
resonances with masses near 2~GeV, a situation that is experimentally 
advantageous.  In addition, baryon resonances have large widths and are often 
overlapping.  Studies of different final states can provide for important 
cross checks in quantitatively understanding the contributing amplitudes.  
Note that although the two ground-state hyperons have the same valence quark 
structure ($uds$), they differ in isospin, such that intermediate $N^*$ resonances 
can decay strongly to $K^+\Lambda$ final states, while both $N^*$ and $\Delta^*$
decays can couple to $K^+\Sigma^0$ final states.

The search for missing resonances requires more than identifying features in 
the mass spectrum. QCD cannot be directly tested with $N^*$ spectra 
without a model for the production dynamics\cite{lee}. The $s$-channel 
contributions are known to be important in the resonance region in order to 
reproduce the invariant mass ($W$) spectra, while $t$-channel meson exchange 
is also necessary to describe the diffractive part of the production and 
$u$-channel diagrams are necessary to describe the backward-going processes.  
Thus measurements that can constrain the phenomenology for these reactions are 
just as important as finding one or more of the missing resonances.  

Theoretically, there has been considerable effort during 
the past decade to develop models for the $KY$ photo- and electroproduction
processes.  However, the present state of understanding is still limited 
by a sparsity of data.  Model fits to the existing cross section data are 
generally obtained at the expense of many free parameters, which leads 
to difficulties in constraining existing theories.  
Moreover, cross section data alone are not sufficiently sensitive to 
fully understand the reaction mechanism, as they probe only a small 
portion of the full response.  In this regard, measurements of spin 
observables are essential for continued theoretical development in this 
field, as they allow for improved understanding of the dynamics of this 
process and provide for strong tests of QCD-inspired models.

In this talk I focus on the strangeness physics program in Hall B
at Jefferson Laboratory using the CLAS detector\cite{mecking}. Presently 
there is very limited knowledge of $N^*,\Delta^* \to KY$ couplings.  With 
the existing CLAS program, the present lack of data 
will be remedied with a wealth of high quality measurements spanning a broad 
kinematic range.

\section{$KY$ Photoproduction}

Photoproduction measurements for $K^+\Lambda$ and $K^+\Sigma^0$ made with
CLAS have provided both differential cross sections and
hyperon polarizations.  The data shown here were collected at electron
beam energies of 2.4 and 3.1~GeV.  This gives rise to measurements
spanning photon energies from threshold $E_\gamma$=0.911~GeV ($W$=1.61~GeV)
up to $E_\gamma$=2.95~GeV ($W$=2.53~GeV).  The final state hyperons were 
reconstructed from the $(\gamma,K^+)$ missing mass.  Detection of the
decay proton from the hyperon was also required.  The average hyperon mass 
resolution was $\sigma$=8.5~MeV.  The $\Lambda$ and $\Sigma^0$ events were 
separated from the pion mis-identification background using lineshape fits to 
the missing mass spectra in each bin of photon energy and kaon angle.

CLAS has already published photoproduction data from another analysis where
only the final state $K^+$ was detected\cite{mcnabb}.  The results shown here
represent a new analysis\cite{bradford} with very different systematics.  The 
two separate analyses are now in good agreement within the associated 
uncertainties, giving us full confidence in the CLAS results.
 
\begin{figure}[htbp]
\vspace{5.5cm}
\includegraphics{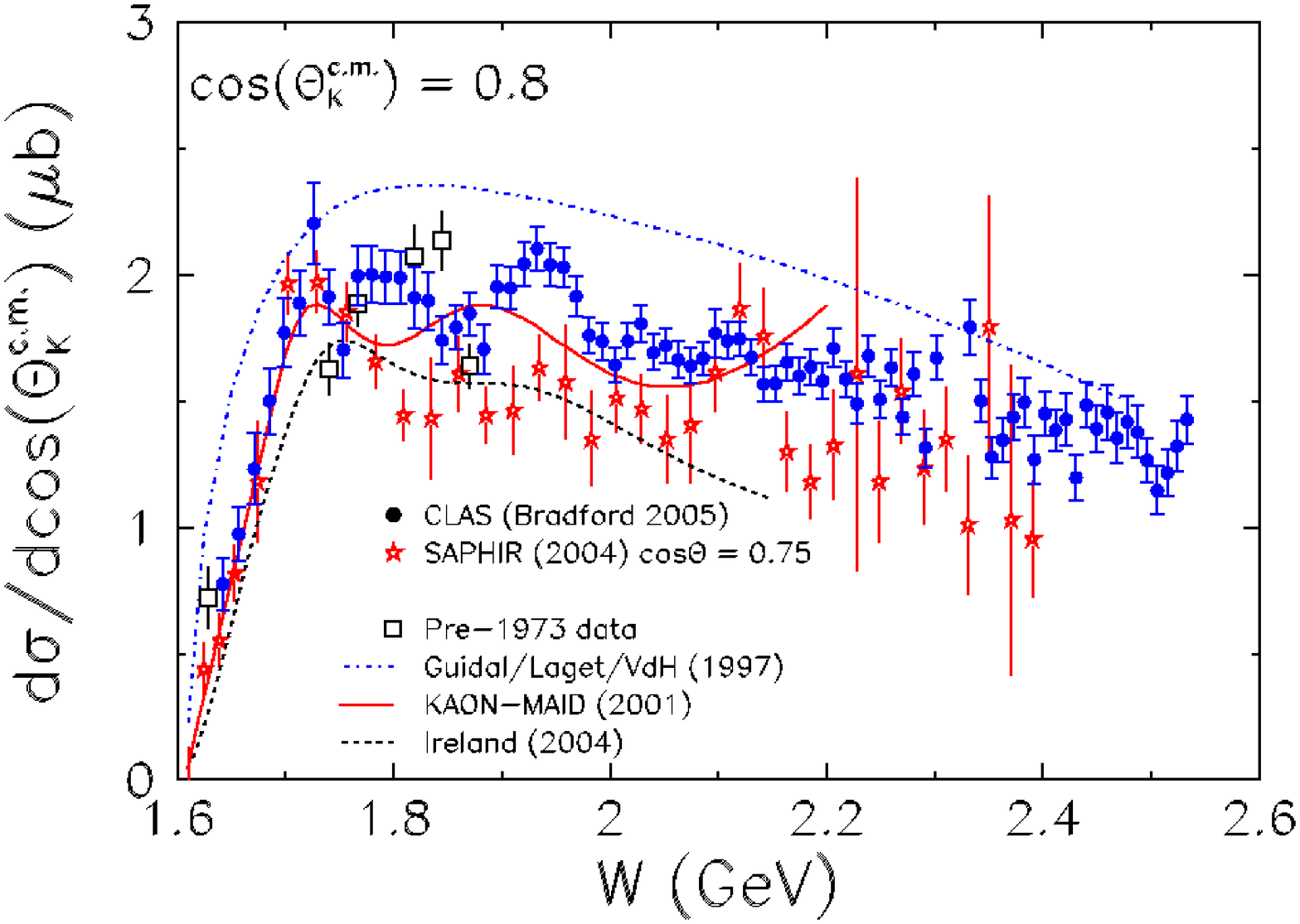}
\includegraphics{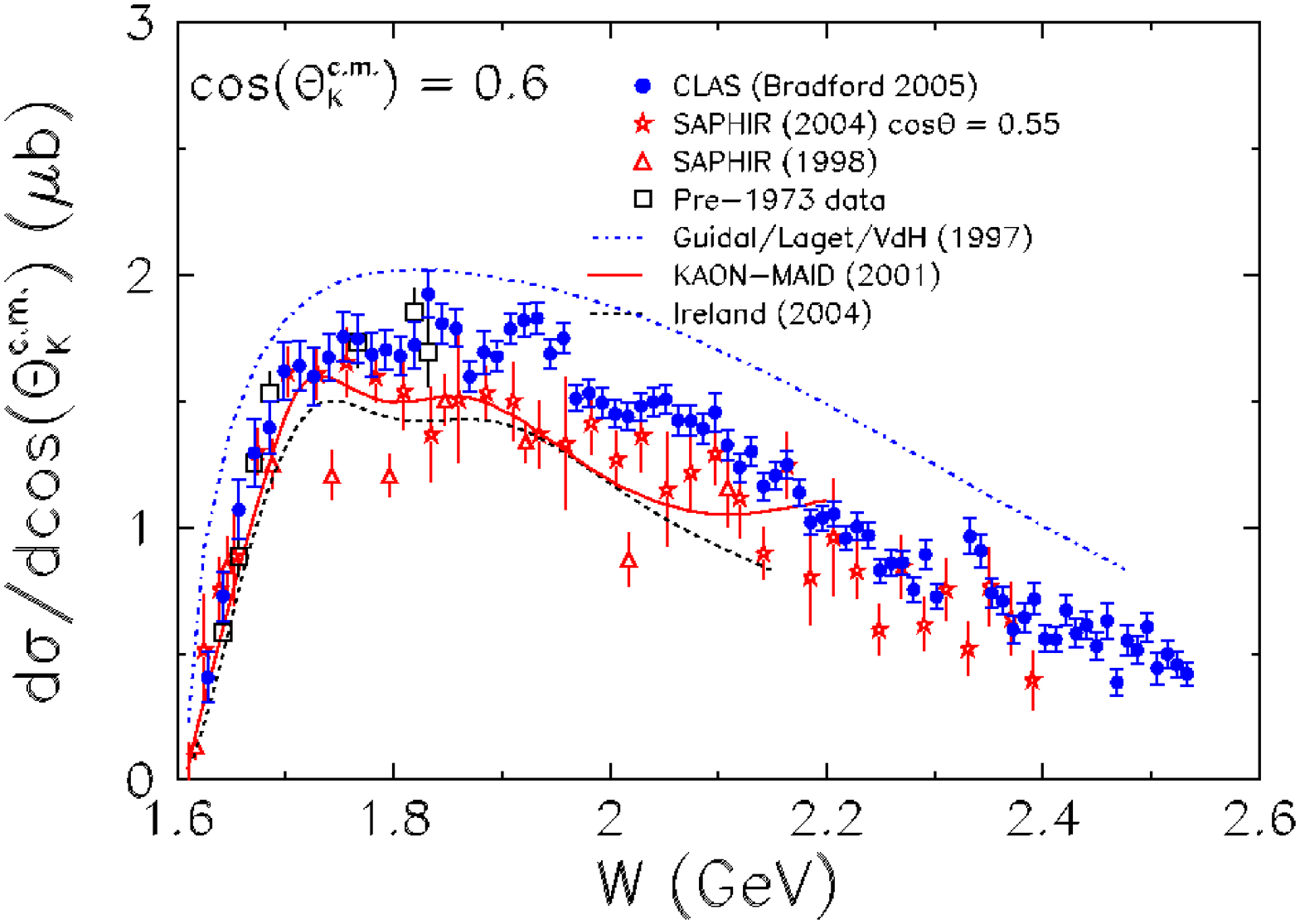}
\includegraphics{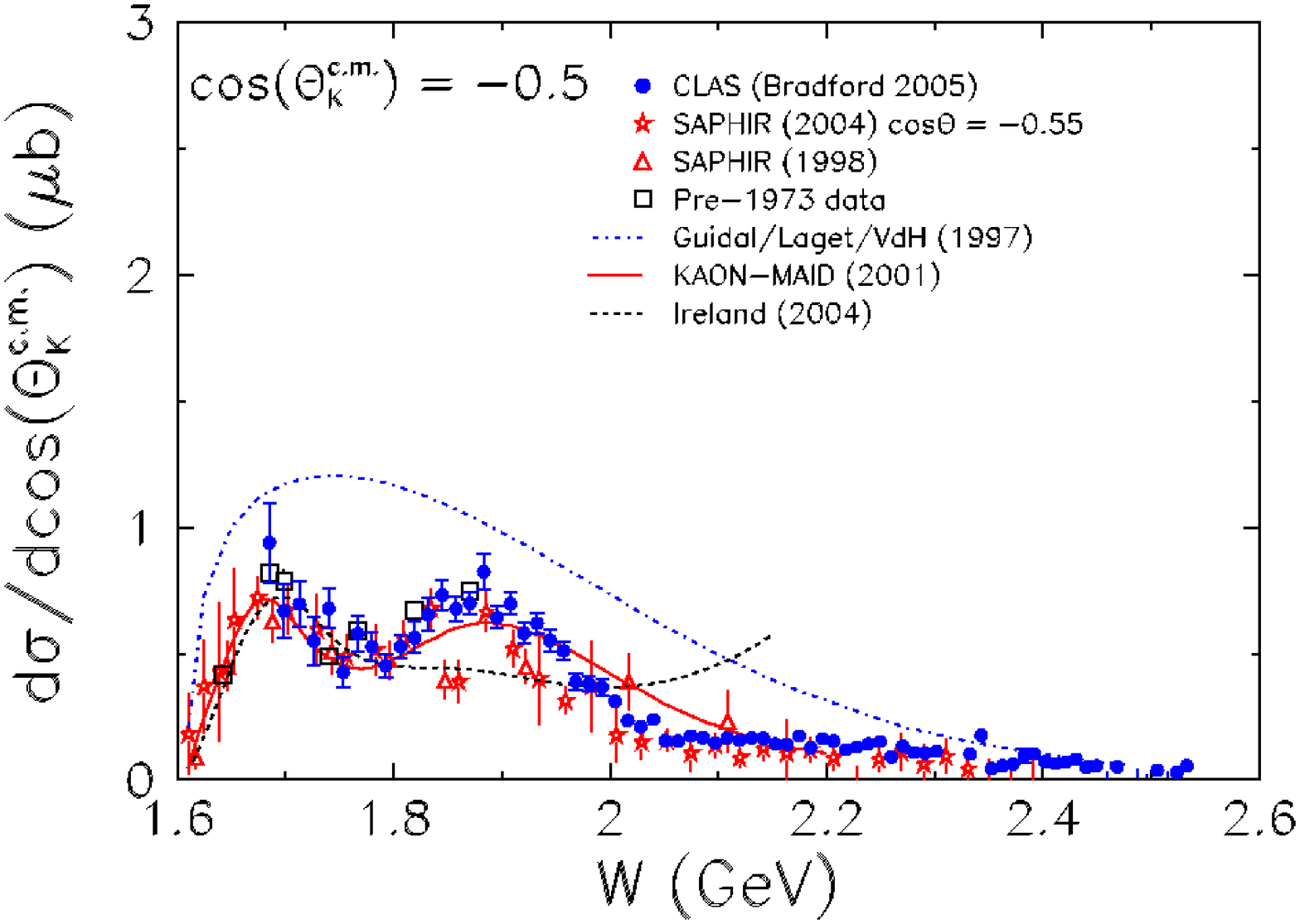}
\includegraphics{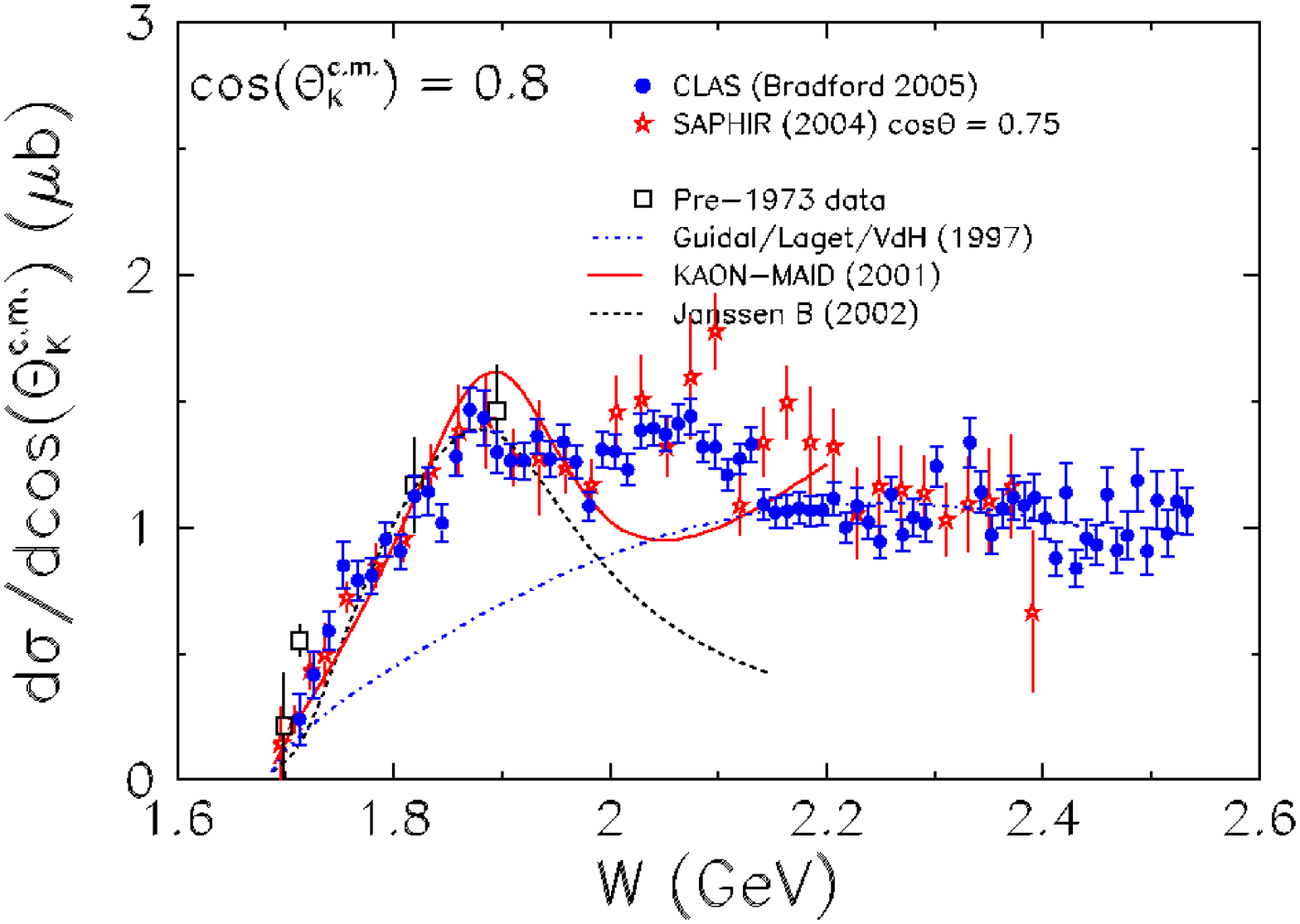}
\includegraphics{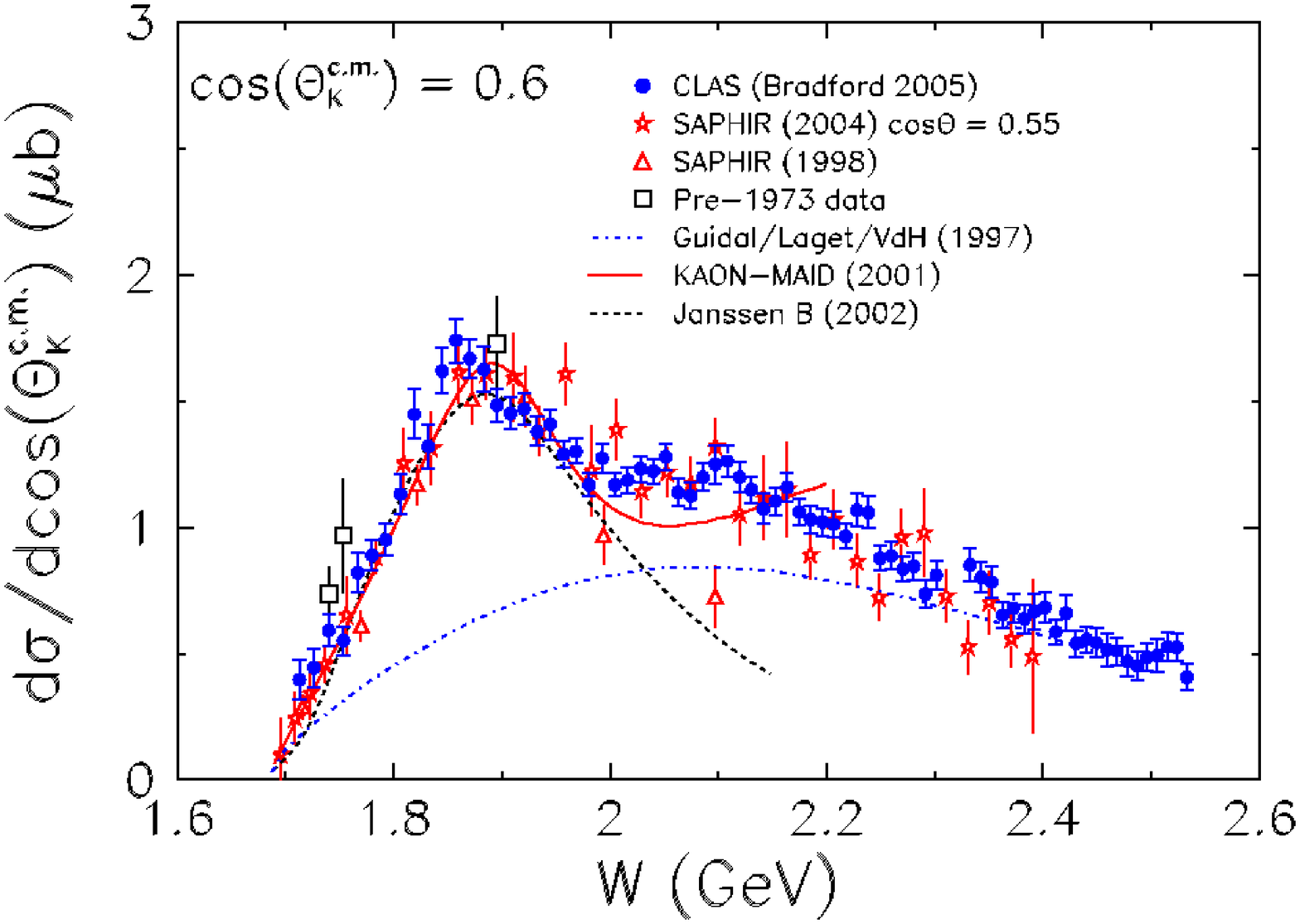}
\includegraphics{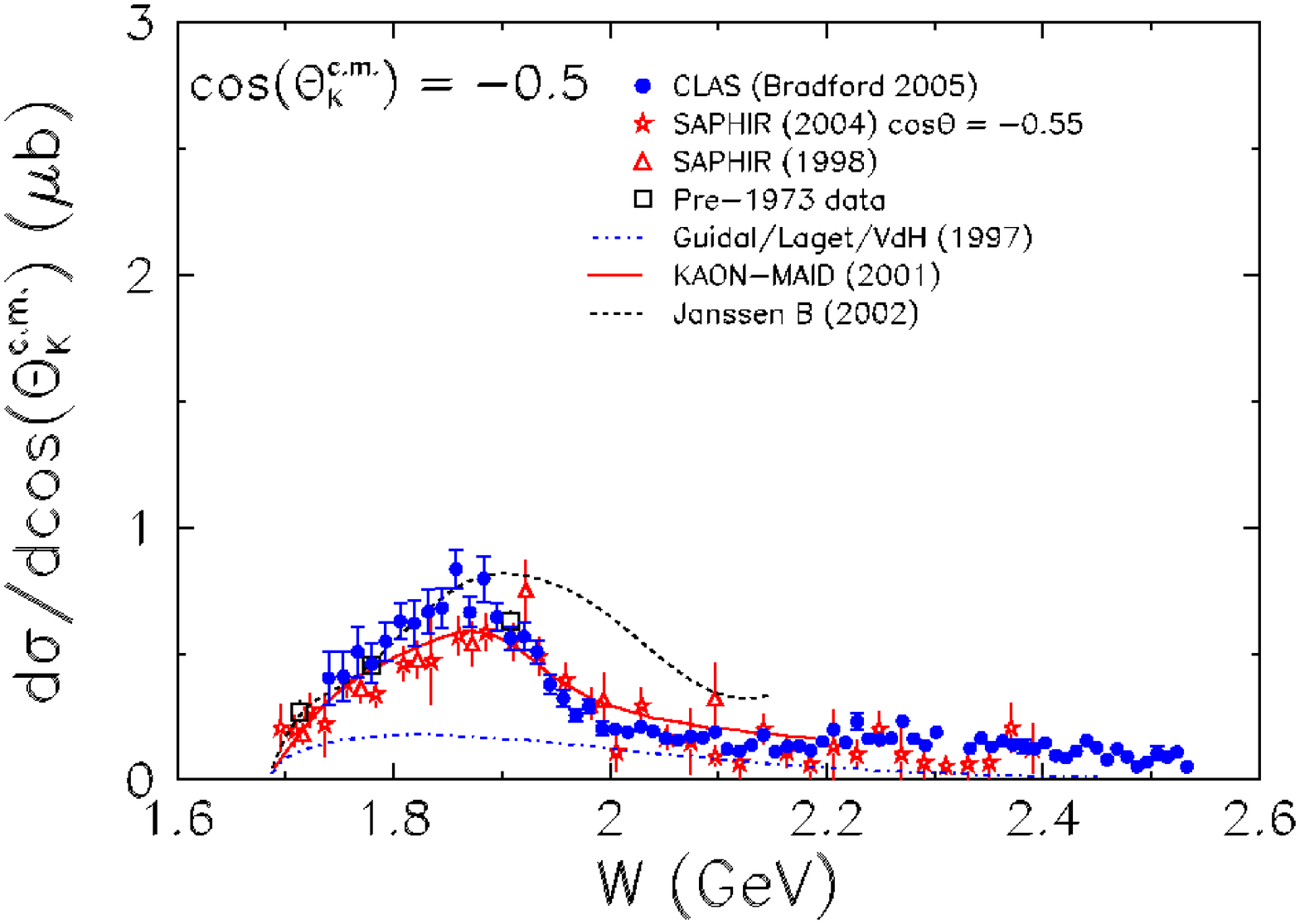}
\caption[]{Differential cross sections for $K^+\Lambda$ (top) and $K^+\Sigma^0$
(bottom) vs. $W$ for three kaon angle bins\cite{bradford} (solid circles). 
Data from Bonn/SAPHIR\cite{saphir} (open triangles) are also shown.  The curves 
are calculations from KAON-MAID\cite{kaonmaid} (solid), Ireland\cite{ireland} 
(dashed), Janssen\cite{janssen} (dashed), and Guidal\cite{guidal} (dot-dashed).}
\label{diffcs}
\end{figure}

Figure~\ref{diffcs} shows a sample of the CLAS differential cross sections for 
$K^+\Lambda$ and $K^+\Sigma^0$ photoproduction as a function of the invariant 
energy $W$ for angle bins at $\cos \theta_K^*$=0.8, 0.6, and -0.5.  The
different angle bins allow us to vary the relative contributions to the
$s$, $t$, and $u$ reaction channels.  Existing data from SAPHIR at 
Bonn\cite{saphir} are also included.  The data are compared with effective 
Lagrangian calculations from Mart/Bennhold\cite{kaonmaid}, Ireland\cite{ireland},
and Janssen\cite{janssen}, which are based on adding the non-resonant Born terms 
with a number of resonances and leaving their coupling constants as free 
parameters bounded loosely by SU(3) predictions.  These models have been 
developed from fits to the Bonn data, however they only reproduce the
threshold region of the data.  Much beyond about 200~MeV above threshold 
the calculations do not reflect the CLAS data.  The data are also compared with a 
Reggeon exchange model\cite{guidal} that uses only $K$ and $K^*$ exchanges, 
with no resonance contributions.  The prediction was made using a model that fit 
higher energy kaon electroproduction data well.

For the $K^+\Lambda$ data the broad structure just above the threshold region 
is typically accounted for by the known $S_{11}$(1650), $P_{11}$(1710), and 
$P_{13}$(1720) resonances.  Centered at roughly 1.9~GeV is another broad 
structure, first seen in the Bonn data, that remains unexplained, whose
position and width vary with kaon angle.  This has been interpreted by Mart 
and Bennhold\cite{mart} as evidence for a missing $D_{13}$(1900) resonance, 
where the assignment was consistent with the measured angular distributions, 
as well as a predicted quark model state\cite{capstick}.  However, other groups 
have shown that the same data can also be explained by accounting for
$u$-channel hyperon exchanges\cite{saghai} or with an additional
$P$-wave resonance\cite{janssen}.  Interestingly, the Regge model fully saturates 
the strength in the reaction, leaving no room for significant $s$ and $u$
channel contributions.

For the $K^+\Sigma^0$ data, there is a single peak in the differential
cross sections at about 1.9~GeV.  This has been associated with
a cluster of $\Delta$ resonances in this mass range.  However
both isospin 1/2 ($N^*$) and isospin 3/2 ($\Delta^*$) resonances
can contribute to this final state.  These data, as well as the
Bonn data, show evidence for resonant decays to $K^+\Sigma^0$.
The Regge model here now provides only a fraction of the reaction
strength with significant contributions possible from the $s$ and
$u$ reaction channels.

\begin{figure}[htbp]
\vspace{4.8cm}
\includegraphics{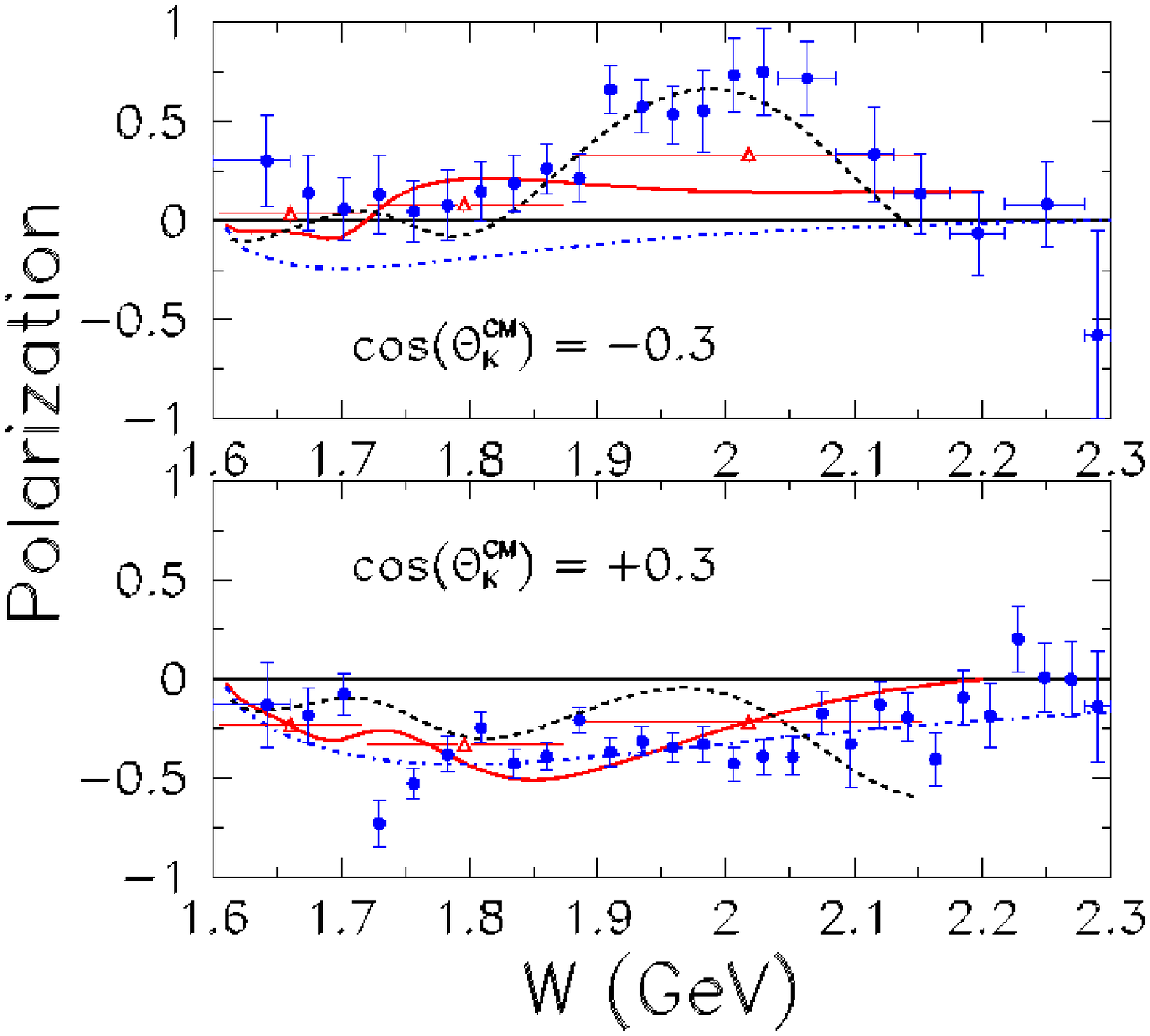}
\includegraphics{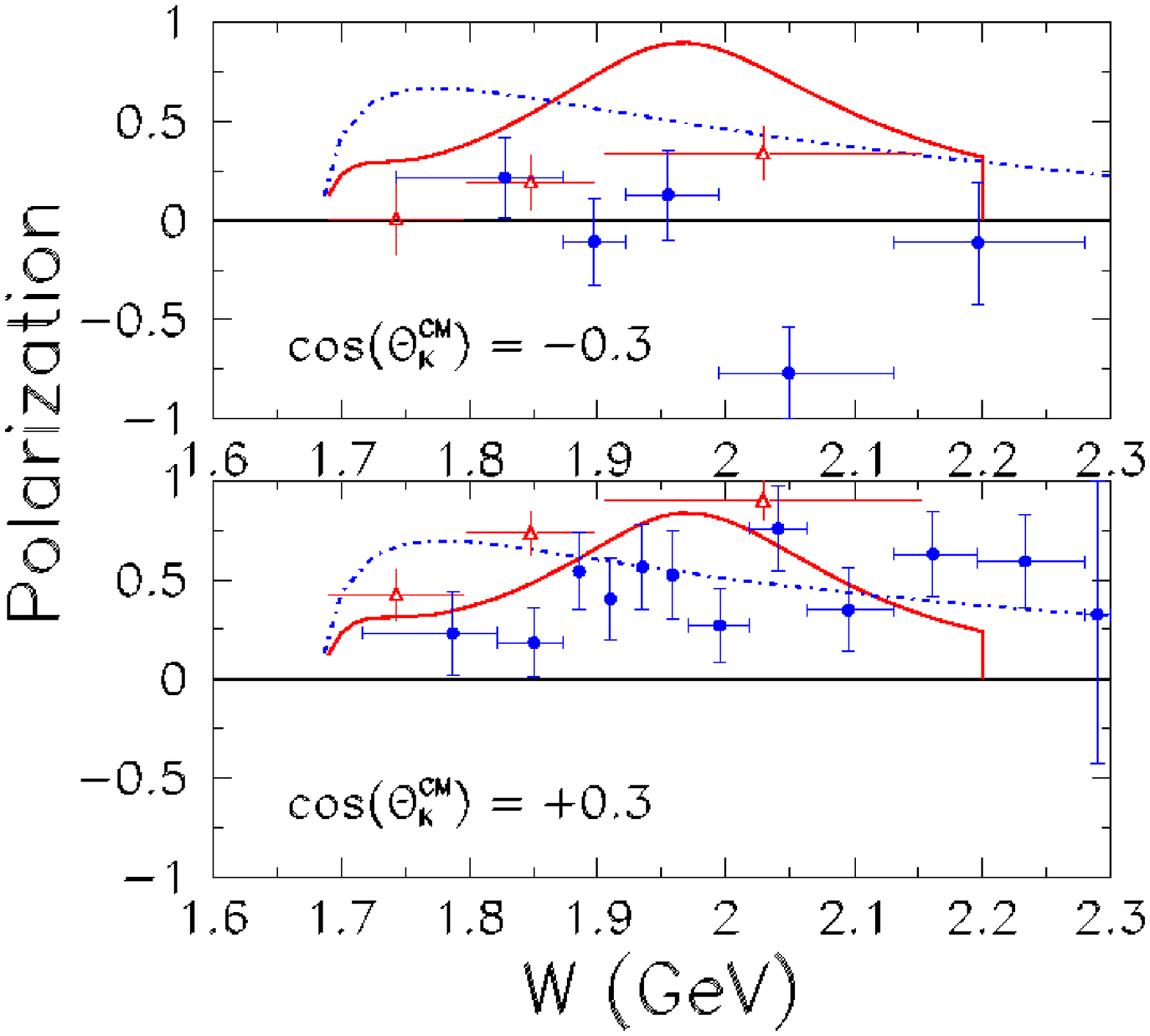}
\caption[]{Induced polarization of the $\Lambda$ and $\Sigma^0$ hyperon 
as a function of $W$ for two kaon angle bins\cite{mcnabb}.  The curves are 
calculations from KAON-MAID\cite{kaonmaid} (solid), Janssen\cite{janssen} 
(dashed), and Guidal\cite{guidal} (dot-dashed).}
\label{gampol}
\end{figure}

Another part of the photoproduction analysis program is to measure
the induced hyperon polarization with an unpolarized beam
and target.  An attractive feature of the hyperon decay is its well 
known self-analyzing nature.  The hyperon polarization is revealed by 
the asymmetry in the angular distribution of the protons from the 
mesonic decay of the hyperon.  From parity conservation, the only allowed 
polarization component is along the axis perpendicular to the $K^+Y$ 
reaction plane.  Measurement of this observable is important since it is 
related to interferences of the imaginary part of resonant amplitudes with 
other amplitudes, including Born terms.  These data are shown in 
Fig.~\ref{gampol} as a function of $W$ for two kaon angle bins at 
$\cos \theta_K^*$=0.3 and -0.3\cite{mcnabb}.  The CLAS data provide the 
first precision data for this observable.  The data show a sizeable negative 
$\Lambda$ polarization for forward-going kaons, and an equally sizeable 
positive polarization when the kaons go backward.  The basic trend is 
reversed in the $\Sigma^0$ data.  The CLAS results are consistent with 
some older data points from Bonn\cite{saphir}.  CLAS has also completed 
measurements of the beam recoil polarization observables $C_x$ and 
$C_z$\cite{bradford_talk}.

Neither the hadrodynamic nor Regge calculations reproduce the magnitudes or 
the trends seen in the hyperon polarization data across the broad kinematic
region covered.  The significant discrepancies between the calculations and 
the data imply that these data can serve to provide for significant new 
constraints on the model parameters.  

A recent coupled-channels analysis\cite{sarantsev} of photoproduction
data from SAPHIR and CLAS, as well as beam asymmetry data from SPring-8/LEPS 
for $K^+\Lambda$\cite{zegers} and data from $\pi$ and $\eta$ photoproduction,
reveals evidence for new baryon resonances in the high $W$ mass region.  The
full suite of data can only be satisfactorily fitted by including a new
$P_{11}$ state at 1840~MeV and two $D_{13}$ states at 1870 and 2130~MeV.
The only $\Delta$ state that contributed significantly to the $K^+\Sigma^0$
final state is the $D_{33}(1940)$.  This analysis has certain ambiguities that 
can be resolved or better constrained by incorporating the expansive set of 
electroproduction data from CLAS.

\section{$KY$ Electroproduction}

CLAS has measured exclusive $K^+\Lambda$ and $K^+ \Sigma^0$ electroproduction 
on the proton for a range of momentum transfer $Q^2$ from 0.5 to 4.5~(GeV/c)$^2$ 
with electron beam energies from 2.6 to 5.7~GeV.  For this talk I will focus 
attention on our 2.6~GeV data set. The final state hyperons were reconstructed
from the $(e,e'K^+)$ missing mass.  The average hyperon resolution was about
8~MeV, similar to what was found for photoproduction.  The hyperon yields
were extracted using Monte Carlo templates with a background determined from
the data associated with pions misidentified as kaons.

The most general form for electroproduction cross section of the kaon from 
an unpolarized-proton target is given by:

\begin{displaymath}
\label{csec1}
\frac{d^4\sigma}{dQ^2 dW d\Omega_K^*} \equiv \sigma_0 = \Gamma_v [ \sigma_T +
\epsilon \sigma_L + \epsilon \sigma_{TT} \cos 2 \Phi +
\sqrt{2 \epsilon (\epsilon + 1)} \sigma_{LT} \cos \Phi ].
\end{displaymath}

\noindent
In this expression, the cross section is decomposed into four structure
functions, $\sigma_T$, $\sigma_L$, $\sigma_{TT}$, and $\sigma_{LT}$, 
which are in general functions of $Q^2$, $W$, and $\theta_K^*$ only.
$\Gamma_v$ represents the virtual photon flux factor, $\epsilon$ is the
virtual photon polarization, and $\Phi$ is the angle between the electron
scattering and hadronic reaction planes.  One of the goals of the
electroproduction program is to provide a detailed tomography of the
structure functions vs. $Q^2$, $W$, and $\cos \theta_K^*$.  In a first
phase of the analysis at CLAS, we have measured the unseparated cross section
($\sigma_U = \sigma_T + \epsilon \sigma_L$) 
and, for the first time in the resonance region away from parallel kinematics, 
the interference cross sections $\sigma_{TT}$ and $\sigma_{LT}$.  At the 
amplitude level, these interference responses are related to real photon 
measurements of the polarized beam asymmetry, and so they are sensitive to some 
of the same structure information.  Exploiting the $\Phi$ dependence of the 
reaction allows us to extract these responses from the CLAS data.
The $Q^2$ dependence of the data provides sensitivity to the associated form
factors.

\begin{figure}[htbp]
\vspace{10.8cm}
\includegraphics{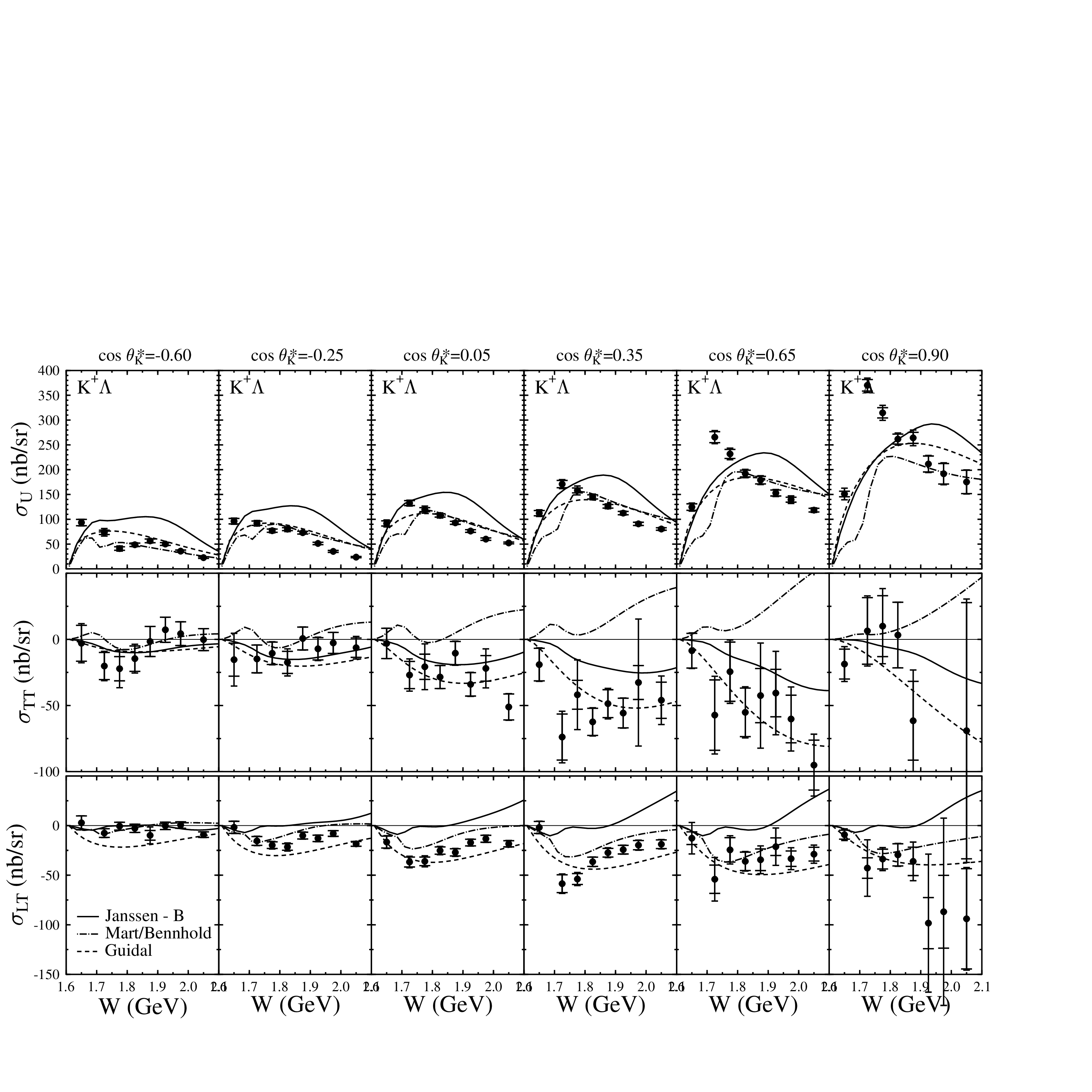}
\includegraphics{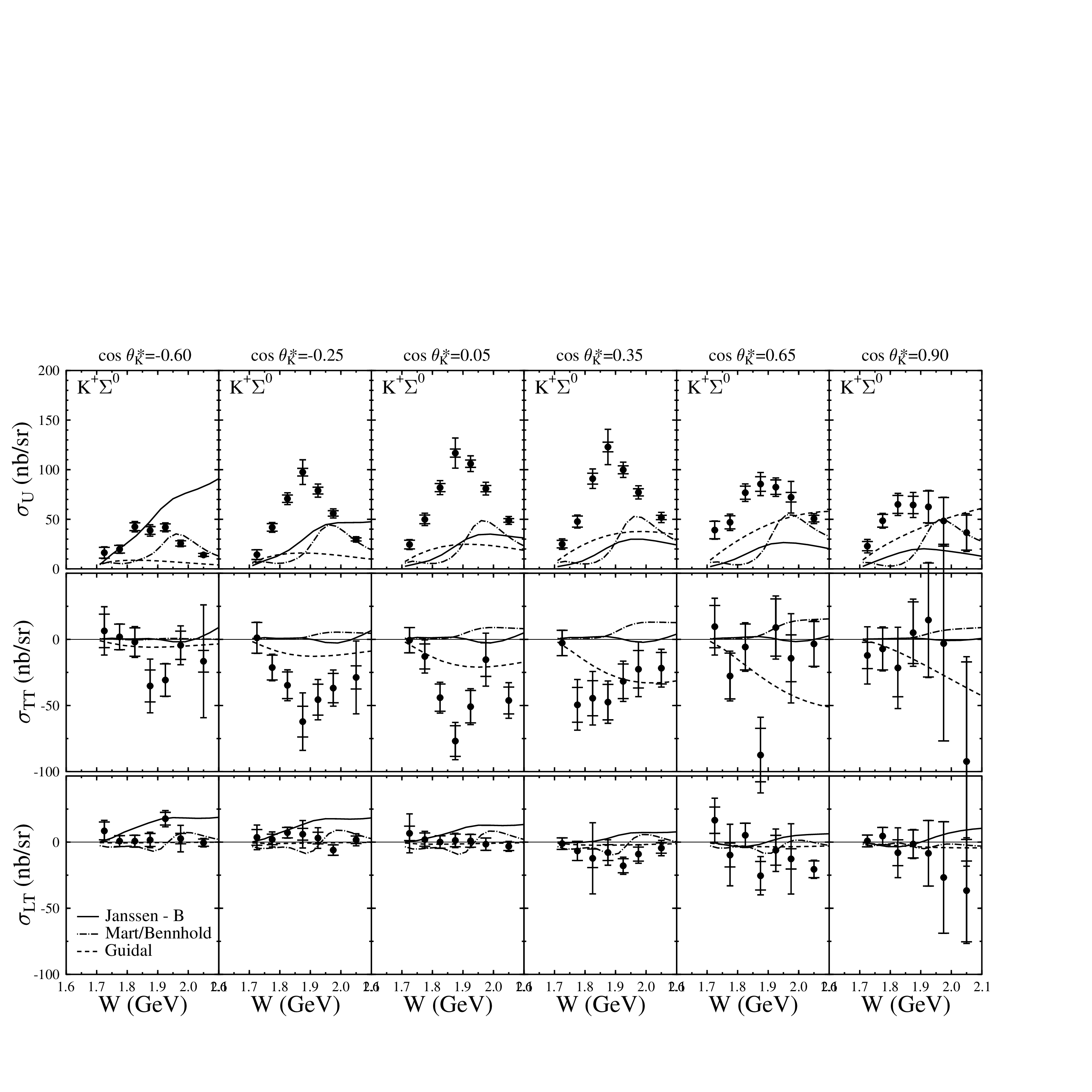}
\caption[]{Preliminary separated structure functions $\sigma_U$, $\sigma_{LT}$, 
and $\sigma_{TT}$ for the $K^+\Lambda$ (top) and $K^+\Sigma^0$ final state
(bottom) at 2.6~GeV and $Q^2$=0.65~(GeV/c)$^2$.  The curves correspond to the 
calculations of Janssen\cite{janssen}, Guidal\cite{guidal}, and
Mart/Bennhold\cite{mart}.}
\label{hyp_plot}
\end{figure}

A small sample of the available results from this analysis is shown in
Fig.~\ref{hyp_plot} vs. $W$ for each of our six angle bins for the kaon.  
The kinematic dependence of the unpolarized structure functions shows that 
$\Lambda$ and $\Sigma^0$ hyperons are produced very differently.  $\sigma_U$ 
at forward angles for $K^+\Lambda$ is dominated by a structure at $W$=1.7~GeV.  
For larger kaon angles, a second structure emerges at about 1.9~GeV, consistent 
with the signature in photoproduction.  $\sigma_{TT}$ and $\sigma_{LT}$ are
clearly non-zero and reflect the structures in $\sigma_U$.  The fact that
$\sigma_{LT}$ is non-zero is indicative of longitudinal strength.  For
the $K^+\Sigma^0$ final state, $\sigma_U$ is centrally peaked, with a 
single broad structure at 1.9~GeV.  This is consistent with the photoproduction
data.  $\sigma_{TT}$ reflects the features of $\sigma_U$, with $\sigma_{LT}$
consistent with zero everywhere, indicative of $\sigma_L$ being consistent
with zero.

To date we have completed analysis of data sets at 2.6 and 4.2~GeV and
have performed a Rosenbluth separation for several $W$ bins over the full 
kaon angular range for a single bin at $Q^2$=1.0~(GeV/c)$^2$ where the data 
sets overlap.  A crucial part in this analysis of extracting absolute
cross sections is to minimize the physics model dependence of the detector
acceptance function, radiative corrections, and bin-centering factors.
We estimate an average absolute systematic uncertainty on these data points
of about 15\%.

\begin{figure}[htbp]
\vspace{4.5cm}
\includegraphics{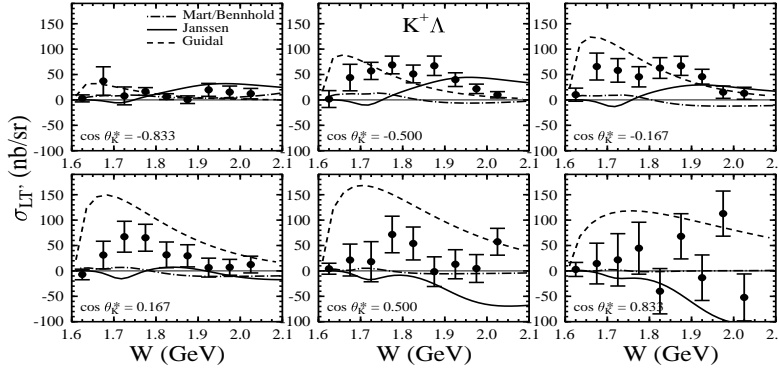}
\caption[]{Preliminary CLAS results\cite{nasser} for the $K^+\Lambda$ structure
function $\sigma_{LT'}$ vs. $W$ for six kaon angle bins for 
$Q^2$=0.65~(GeV/c)$^2$. The curves are from the model calculations 
from Janssen\cite{janssen}, Guidal\cite{guidal}, and Mart/Bennhold\cite{mart}.}
\label{fifth}
\end{figure}

The polarized-beam asymmetry provides access to the fifth structure function 
$\sigma_{LT'}$. This observable probes imaginary parts of the interfering 
$L$ and $T$ amplitudes (as opposed to the real parts of the interference from 
$\sigma_{LT}$).  These imaginary parts vanish identically if the resonant state 
is determined by a single complex phase, which is the case for an isolated 
resonance.  A representative sample of our data at 2.6~GeV and 
$Q^2$=0.65~(GeV/c)$^2$ is shown in Fig.~\ref{fifth} for the $K^+\Lambda$ 
final state\cite{nasser}.  The calculations shown are not able to
reproduce the features seen in the data.

\begin{figure}[htbp]
\vspace{5.0cm}
\includegraphics{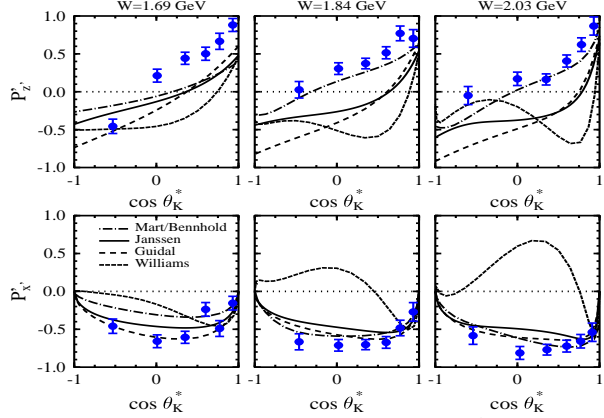}
\caption[]{CLAS transferred polarization\cite{carman03} from $\vec{e} p \rightarrow 
e'K^+ \vec{\Lambda}$ vs. $\cos \theta_K^*$ at 2.6~GeV for three different 
$W$ bins summed over $\Phi$ and $Q^2$.   The curves are for 
effective Lagrangian calculations of Janssen\cite{janssen}, Guidal\cite{guidal},
Bennhold\cite{haber}, and WJC\cite{wjc92}.}
\label{dpol}
\end{figure}

The first measurements of spin transfer from a longitudinally polarized 
electron beam to the $\Lambda$ hyperon produced in the exclusive 
$p(\vec{e},e'K^+)\vec{\Lambda}$ reaction have recently been completed at 
CLAS\cite{carman03}.  A sample of the results 
highlighting the angular dependence of $P'$ summed over all $Q^2$ for
three different $W$ bins is shown in Fig.~\ref{dpol} at 2.6~GeV.  
The polarization along the virtual photon direction $P_{z'}'$ decreases 
with increasing $\theta_K^*$, while the orthogonal component in the hadronic
reaction plane $P_{x'}'$ is constrained to be zero at $\cos \theta_K^*$ = 
$\pm$1 due to angular momentum conservation, and reaches a minimum at 
$\theta_K^* \sim 90^{\circ}$.  The component normal to the hadronic
reaction plane $P_{y'}'$ is statistically consistent with zero as 
expected.

The transferred polarization data are compared with three effective
Lagrangian models that include a different subset of resonances.  It
is interesting that the model with the best agreement includes the
$D_{13}$(1900) resonance\cite{haber}.  The accuracy of the measurements, 
coupled with the spread in the theory predictions, clearly indicates that 
these data are sensitive to the resonant and non-resonant structure of the 
intermediate state.

\begin{figure}[htbp]
\vspace{3.6cm}
\includegraphics{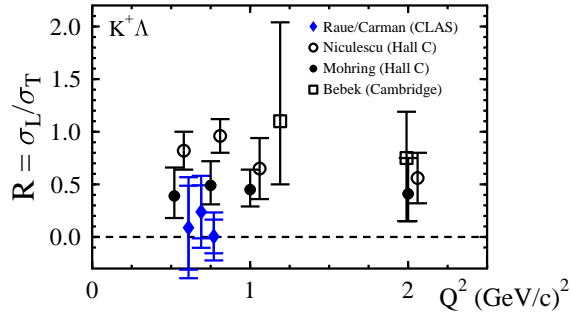}
\caption[]{Ratio of longitudinal to transverse structure functions vs. $Q^2$. 
The Niculescu results\cite{Niculescu}, which were superseded by the
Mohring results\cite{Mohring03}, are offset in $Q^2$ for clarity.}
\label{dpol_rat}
\end{figure}

The transferred polarization data have also been used to measure the ratio
$\mathcal{R}=\sigma_L/\sigma_T$\cite{raue}.  This can be done by extrapolating 
the $P'$ data to $\theta_K^*$=0$^{\circ}$, where $\mathcal{R} = 
(1/\epsilon)(c_0/P'_{z'} - 1)$. Here $c_0$ represents a kinematic factor.  
This method provides a complementary approach from the standard Rosenbluth 
measurement in a situation with different systematics.  Existing data from Hall C 
have remained controversial.  The new results, shown in Fig.~\ref{dpol_rat}, are 
consistent with, although systematically smaller than, the latest Hall C results.  
They indicate that $\mathcal{R}$ is reasonably constant with $Q^2$ with small 
values for $\sigma_L$.

\section{Summary and Conclusions}

In this talk I have reviewed some of the key reasons why the photo-
and electroproduction processes of open-strangeness production are 
important for the investigation of baryonic structure and missing 
quark model states.  I have discussed several aspects of the 
CLAS strangeness physics program highlighting the breadth and quality 
of our data sets.  Results were presented for cross sections and
single and double-polarization observables for $K^+\Lambda$ and
$K^+\Sigma^0$ photo- and electroproduction.  Our analyses indicate that 
the data are highly sensitive to the ingredients of the models, including 
the specific baryonic resonances included, along with their associated 
form factors and coupling constants.  The production dynamics for
$K^+\Lambda$ and $K^+\Sigma^0$ are also seen to be very different.

Work on publication of the full set of hyperon cross section and polarization 
data sets reported here is in progress.  The main 
qualitative conclusion seems clear: these data show significant 
unexplained baryon resonance structure at higher masses.  While the 
comparison of the effective Lagrangian calculations to the data is 
illustrative to highlight the present deficiencies in the current models 
and their parameter values, the next step in the study of the reaction 
mechanism is to include our data in the available data base and to refit 
the set of coupling strengths.  Additionally new amplitude-level
analyses are called for to more fully unravel the contributions to
the intermediate state.

This work has been supported by the U.S. Department of Energy and the 
National Science Foundation.


\begin{thebibliography}{99}

\bibitem{capstick}
S. Capstick and W. Roberts, Phys. Rev. D {\bf 58}, 74011 (1998).

\bibitem{lee}
T.-S.H. Lee and T. Sato, Proceedings of the N*2000 Conference, eds. Burkert
{\it et al.}, (World Scientific, Singapore, 2001), p. 215.

\bibitem{mecking}
B.A. Mecking {\it et al.} {\it (CLAS Collaboration)}, Nucl. Inst. and Meth. A 
{\bf 503}, 513 (2003).

\bibitem{mcnabb}
J. McNabb {\it et al.} {\it (CLAS Collaboration)}, Phys. Rev. C {\bf 69}, 042201 
(R) (2004).

\bibitem{bradford}
R. Bradford {\it et al.} {\it (CLAS Collaboration)}, nucl-ex/0509033, 
submitted to Phys. Rev. C, (2005).

\bibitem{saphir}
M.Q. Tran {\it et al.}, Phys. Lett B {\bf 445}, 20 (1998).

\bibitem{kaonmaid}
T. Mart {\it et al}, ``KaonMAID 2000'' at www.kph.uni-mainz.de/MAID.

\bibitem{ireland}
D.G. Ireland, S. Janssen, and J. Ryckebusch, Nucl. Phys. A {\bf 740}, 147 (2004).

\bibitem{janssen}
S. Janssen {\it et al.}, Phys. Rev. C 65, 015201 (2002).

\bibitem{guidal}
M. Guidal, J.M. Laget, and M. Vanderhaegen, Nucl. Phys. {\bf A627},
645 (1997).

\bibitem{mart}
T. Mart and C. Bennhold, Phys. Rev. C {\bf 61}, 012201 (2000).

\bibitem{saghai}
B. Saghai, nucl-th/0105001, (2001). 

\bibitem{bradford_talk}
R. Bradford, see talk in these proceedings.

\bibitem{sarantsev}
A.V. Sarantsev {\it et al.}, hep-ex/0506011, submitted to Eur. Phys. J, (2005).

\bibitem{zegers} 
R.G.T. Zegers {\it et al.} {\it (LEPS Collaboration)}, Phys. Rev. Lett. {\bf 91},
092001 (2003).

\bibitem{nasser}
R. Nasseripour {\it et al.} {\it (CLAS Collaboration)}, to be submitted for
publication.

\bibitem{haber}
H. Haberzettl {\it et al.}, Phys. Rev. C {\bf 58}, R40 (1998).

\bibitem{wjc92}
R.A. Williams, C. Ji, and S.R. Cotanch, Phys. Rev. C {\bf 46}, 1617 (1992).

\bibitem{carman03}
D.S. Carman {\it et al.} {\it (CLAS Collaboration)}, Phys. Rev. Lett. {\bf 90}, 
131804 (2003).

\bibitem{raue}
B.A. Raue and D.S. Carman, Phys. Rev. C {\bf 71}, 065209 (2005).

\bibitem{Niculescu}
G. Niculescu {\it et al.}, Phys. Rev. Lett. {\bf 81}, 1805 (1998).

\bibitem{Mohring03}
R.M. Mohring {\it et al.}, Phys. Rev. C {\bf 67}, 055205 (2003).

\end{thebibliography}
\end{document}